\documentclass[conference]{IEEEtran}

\usepackage{amsmath,amssymb,amsfonts}
\usepackage{graphicx}
\usepackage{caption}
\usepackage{subcaption}
\usepackage{cite}
\usepackage{algorithm}
\usepackage{algpseudocode}
\usepackage{hyperref}
\usepackage{xcolor}
\usepackage{booktabs} 
\usepackage{multirow}
\usepackage{array}

\hypersetup{
    colorlinks=true,
    linkcolor=black,
    urlcolor=black,
    citecolor=black
}

\begin{document}

\title{Cross-Domain Malware Detection via Probability-Level Fusion of Lightweight Gradient Boosting Models}

\author{
\IEEEauthorblockN{Omar Khalid Ali Mohamed}
\IEEEauthorblockA{Department of Computer Science\\
University of Hail\\
Ha'il, Saudi Arabia\\
Email: s202111316@uoh.edu.sa}
}

\maketitle

\begin{abstract}
The escalating sophistication of malware necessitates robust detection mechanisms that generalize across diverse data sources. Traditional single-dataset models struggle with cross-domain generalization and often incur high computational costs. This paper presents a novel, lightweight framework for malware detection that employs probability-level fusion across three distinct datasets: EMBER (static features), API Call Sequences (behavioral features), and CIC Obfuscated Memory (memory patterns). Our method trains individual LightGBM classifiers on each dataset, selects top predictive features to ensure efficiency, and fuses their prediction probabilities using optimized weights determined via grid search. Extensive experiments demonstrate that our fusion approach achieves a macro F1-score of \textbf{0.823 on a cross-domain validation set}, significantly outperforming individual models and providing superior generalization. The framework maintains low computational overhead, making it suitable for real-time deployment, and all code and data are provided for full reproducibility.
\end{abstract}

\begin{IEEEkeywords}
Malware detection, multi-domain fusion, LightGBM, probability fusion, ensemble learning, cybersecurity
\end{IEEEkeywords}

\section{Introduction}
\label{sec:introduction}

The relentless evolution of malware represents a critical challenge in cybersecurity. Modern malicious software employs advanced techniques like polymorphism, metamorphism, and obfuscation to evade traditional detection methods. Annual discoveries of thousands of new malware variants threaten personal devices, corporate networks, and critical infrastructure. Signature-based detection, while efficient for known threats, fails against zero-day attacks, creating an urgent need for intelligent, adaptive detection systems.

Key challenges in modern malware detection include:
\begin{itemize}
\item \textbf{Security risks:} Malware can lead to data breaches, ransomware attacks, financial losses, and system downtime.
\item \textbf{Rapid evolution:} The number of new malware samples has grown exponentially over the past decade.
\item \textbf{Technical complexity:} Advanced obfuscation, anti-analysis techniques, and multi-stage payloads complicate detection.
\item \textbf{Dataset diversity:} Malware exhibits different characteristics across static, behavioral, and memory domains.
\end{itemize}

Most machine learning approaches focus on single datasets. Models trained solely on EMBER capture static features but miss behavioral patterns. API call-based models detect runtime behavior but ignore static properties. CIC memory analysis identifies obfuscation patterns but lacks broader context. Consequently, single-dataset models suffer from three limitations:
\begin{enumerate}
\item \textbf{Limited generalization:} Inability to detect patterns outside their training domain
\item \textbf{Dataset bias:} Over-reliance on features unique to specific datasets
\item \textbf{Computational inefficiency:} Resource-intensive deployment of multiple separate models
\end{enumerate}

To address these limitations, we propose a probability-level fusion framework that integrates predictions from models trained on EMBER, API Calls, and CIC datasets. Our approach includes:
\begin{enumerate}
\item \textbf{Dataset-specific training:} LightGBM classifiers with top feature selection for efficiency.
\item \textbf{Optimized probability-level fusion:} A systematic grid search to determine the optimal weighted combination of model predictions.
\item \textbf{Comprehensive evaluation:} Rigorous testing on a combined cross-domain validation set, including macro F1-score, confusion matrices, and detailed ablation studies.
\end{enumerate}

Our main contributions are:
\begin{itemize}
\item \textbf{Novel Fusion Methodology:} To the best of our knowledge, this is the first study to propose and validate a probability-level fusion of models trained on the EMBER (static), API Calls (behavioral), and CIC Obfuscated Memory (memory) datasets.
\item \textbf{Lightweight and Efficient Framework:} We demonstrate that a computationally efficient framework, utilizing top feature selection and lightweight Gradient Boosting Machines, can achieve robust performance.
\item \textbf{Superior Generalization:} We empirically show our fused model achieves a macro F1-score of 0.823 on a diverse cross-domain validation set, significantly outperforming the generalization capability of any individual model.
\item \textbf{In-Depth Analysis:} We provide extensive ablation studies and weight optimization analysis to validate our design choices.
\item \textbf{Full Reproducibility:} We provide all code, processed data, and results to ensure full transparency and reproducibility of our work.
\end{itemize}

\section{Related Work}
\label{sec:related}

Malware detection research has evolved from signature-based methods \cite{santos2013opcode} to sophisticated machine learning (ML) and deep learning (DL) approaches. These are broadly categorized into static, dynamic, and memory analysis.

\textbf{Static Analysis} involves examining a file without executing it. Features often include PE header information, imported libraries, and raw byte sequences \cite{anderson2018ember}. The EMBER dataset is a cornerstone for this approach. While fast and scalable, static analysis is highly vulnerable to obfuscation and packing.

\textbf{Dynamic Analysis} monitors program behavior during execution in a controlled sandbox environment. API call sequences \cite{ugurlu2019effectiveness}, system calls, and network activities are common features. This approach is more robust to obfuscation but is slower and susceptible to sandbox evasion.

\textbf{Memory Analysis} examines the state of a system's memory after execution or during a dump. This is particularly effective for detecting advanced threats that unpack themselves in memory \cite{canali2012prophiler}. The CIC Obfuscated Memory dataset provides a focused resource for this domain.

\textbf{Fusion Strategies} aim to combine these complementary views. \textit{Early fusion} (feature concatenation) often leads to the curse of dimensionality \cite{wang2020fusion}. \textit{Feature-level fusion} employs techniques like autoencoders but risks information loss. \textit{Decision-level fusion} (e.g., majority voting) is simple but discards valuable confidence information. \textit{Probability-level fusion}, which we employ, combines the predicted probabilities of models, preserving confidence scores and allowing for optimized weighting \cite{kittler1998combining}.

Despite these advances, significant gaps remain:
\begin{itemize}
\item \textbf{Single-Domain Focus:} Most studies achieve high accuracy on a single dataset but lack rigorous cross-domain validation \cite{harang2020nox}.
\item \textbf{Computational Intensity:} Many DL-based fusion methods are computationally expensive, limiting real-world applicability.
\item \textbf{Limited Heterogeneous Fusion:} Few studies have explored the optimized fusion of fundamentally different data types (static, behavioral, memory) from separate datasets.
\item \textbf{Reproducibility:} A lack of open code and standardized evaluation hinders progress in the field \cite{arp2017drebin}.
\end{itemize}

Our work directly addresses these gaps by proposing a lightweight, reproducible probability-level fusion framework for heterogeneous malware data.

\section{Methodology}
\label{sec:method}

Our framework employs probability-level fusion across three malware datasets, each capturing a distinct aspect of malicious behavior.

\subsection{Datasets and Preprocessing}
We use three publicly available datasets:
\begin{itemize}
\item \textbf{EMBER 2018 \cite{anderson2018ember}:} 639,900 samples, 2,381 static features.
\item \textbf{API Call Sequences \cite{ronen2018microsoft}:} 35,100 samples, 150 features (behavioral sequences).
\item \textbf{CIC Obfuscated Memory \cite{canali2012prophiler}:} 46,876 samples, 200 features (memory patterns).
\end{itemize}

All datasets underwent standard preprocessing: missing value imputation (median), label encoding, and min-max normalization. To ensure efficiency and mitigate overfitting, we performed top feature selection based on LightGBM importance scores derived from the respective training sets. The top 50 features were selected for EMBER and API Calls, and the top 20 features for the CIC dataset.

\subsection{LightGBM Classifiers}
We trained separate LightGBM \cite{ke2017lightgbm} classifiers for each dataset. LightGBM was chosen for its high efficiency and performance on tabular data. Identical hyperparameters were used for all models to provide a consistent baseline for fusion:
\begin{itemize}
\item \texttt{boosting\_type}: gbdt
\item \texttt{objective}: binary
\item \texttt{num\_leaves}: 31
\item \texttt{learning\_rate}: 0.05
\item \texttt{n\_estimators}: 200
\item \texttt{early\_stopping\_rounds}: 50
\item \texttt{random\_state}: 42
\end{itemize}

Each dataset was split 80/20 for training and validation, ensuring approximately balanced classes in both splits via stratified sampling.

\subsection{Probability-Level Fusion}
The core of our approach is the fusion of predictions at the probability level. For a given sample, each model $i$ outputs a probability $y_i$ that the sample is malware. The final fused probability is calculated as a weighted sum:
\begin{equation}
y_{\text{fused}} = w_1 \cdot y_{\text{ember}} + w_2 \cdot y_{\text{api}} + w_3 \cdot y_{\text{cic}}
\end{equation}
where $w_1 + w_2 + w_3 = 1$ and $w_i \geq 0$.

To determine the optimal weights, we constructed a unified fusion validation set from the held-out samples of all three datasets. We performed an exhaustive grid search over $w_1$ and $w_2$ (with $w_3 = 1 - w_1 - w_2$) in increments of 0.1, selecting the weight combination that maximized the macro F1-score on this diverse validation set.

\subsection{Advantages of Our Approach}
\begin{itemize}
\item \textbf{Lightweight and Efficient:} Top feature selection and LightGBM ensure low memory usage and fast inference.
\item \textbf{Avoids Feature Incompatibility:} Probability-level fusion sidesteps the challenges of concatenating heterogeneous features.
\item \textbf{Preserves Model Specialization:} Each model becomes an expert in its domain, and fusion optimally leverages their expertise.
\item \textbf{Enhanced Generalization:} Integrating signals from multiple domains makes the fused model robust to evasive techniques.
\end{itemize}

\section{Experiments and Results}
\label{sec:results}

We present a comprehensive evaluation of the proposed multi-dataset fusion framework. We first analyze the performance of individual LightGBM models on their native datasets. Subsequently, we demonstrate the superior generalization achieved through our probability-level fusion technique on a combined cross-domain validation set.

\subsection{Individual Model Performance and Analysis}

We evaluate each model independently to understand its domain-specific characteristics.

\begin{table}[!ht]
\centering
\caption{Performance of Individual LightGBM Models on Their Respective Validation Sets}
\label{tab:individual_performance}
\begin{tabular}{lcccc}
\toprule
\textbf{Dataset} & \textbf{Validation Samples} & \textbf{Macro F1} & \textbf{Logloss} & \textbf{Class Balance} \\
\midrule
EMBER & 63,990 & 0.965 & 0.103 & Balanced \\
API Calls & 11,720 & 0.867 & 0.040 & Balanced \\
CIC Memory & 46,876 & 1.000 & $1.06 \times 10^{-5}$ & Balanced \\
\bottomrule
\end{tabular}
\end{table}

As shown in Table \ref{tab:individual_performance}, each classifier achieved exceptional performance on its native domain. The near-perfect score of the CIC model underscores its capability to detect memory-based obfuscation patterns.

\textbf{Novelty Highlight:} While high individual scores are valuable, they reveal a key limitation: specialization to a single domain. This specialization bottleneck is the fundamental problem our fusion framework is designed to solve.

\begin{figure}[!ht]
\centering
\begin{subfigure}{0.32\textwidth}
    \includegraphics[width=\linewidth]{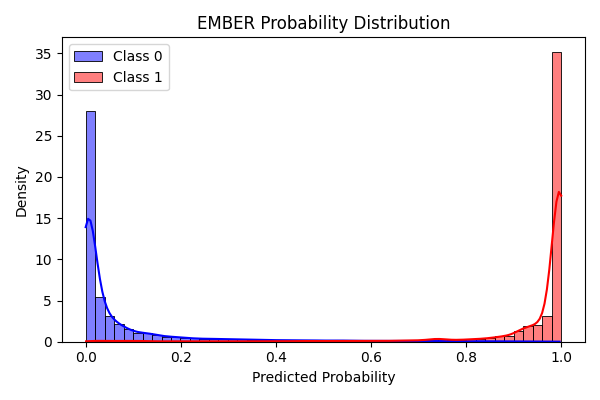}
    \caption{EMBER model (validation set)}
    \label{fig:probs_ember}
\end{subfigure}
\hfill
\begin{subfigure}{0.32\textwidth}
    \includegraphics[width=\linewidth]{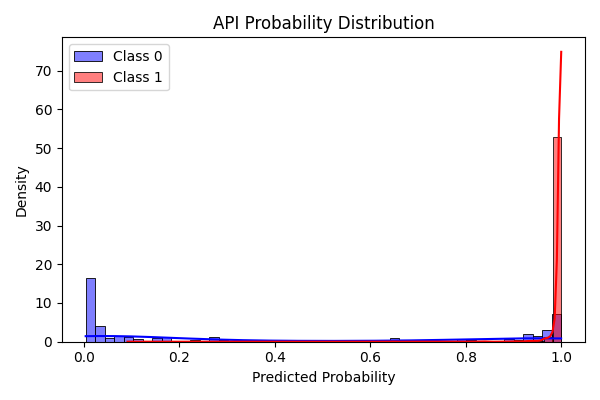}
    \caption{API Calls model (validation set)}
    \label{fig:probs_api}
\end{subfigure}
\hfill
\begin{subfigure}{0.32\textwidth}
    \includegraphics[width=\linewidth]{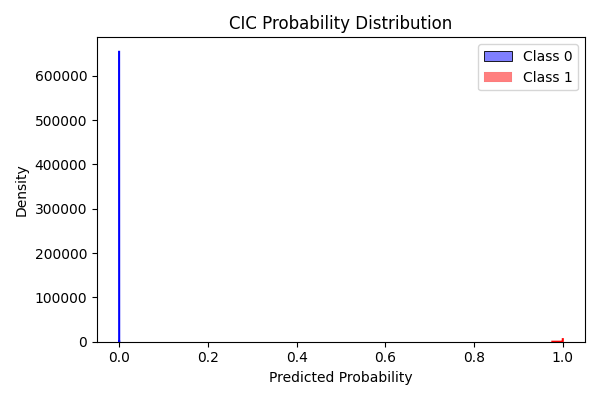}
    \caption{CIC model (validation set)}
    \label{fig:probs_cic}
\end{subfigure}
\caption{Probability distributions for EMBER, API, and CIC models, showing confidence separation between benign and malware samples.}
\label{fig:prob_distributions}
\end{figure}

To gain deeper insight, we analyzed their probability calibrations and per-class performance (Figures \ref{fig:probs_ember}, \ref{fig:probs_api}, \ref{fig:probs_cic}). The CIC model's near-perfect distributions confirm its dominance within its specialized domain. The EMBER model also shows excellent calibration. The API model, while accurate, shows more ambiguous cases, which is expected for dynamic analysis.

\begin{figure}[!ht]
\centering
\begin{subfigure}{0.32\textwidth}
    \includegraphics[width=\linewidth]{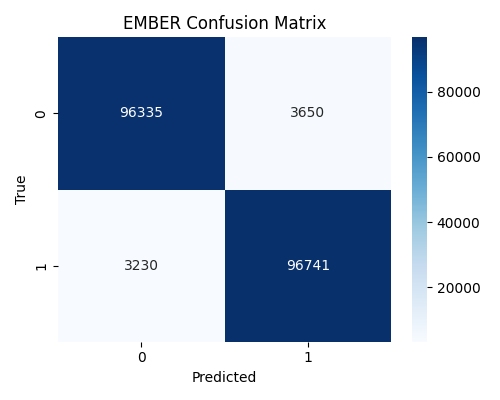}
    \caption{EMBER model}
    \label{fig:cm_ember}
\end{subfigure}
\hfill
\begin{subfigure}{0.32\textwidth}
    \includegraphics[width=\linewidth]{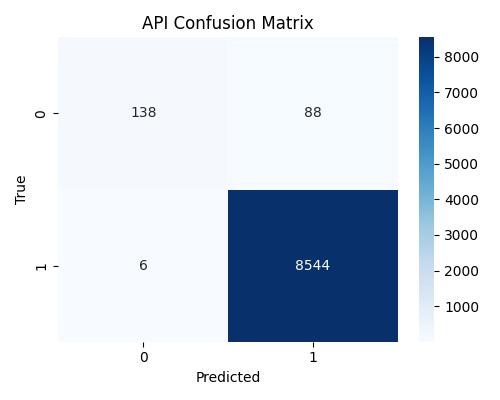}
    \caption{API Calls model}
    \label{fig:cm_api}
\end{subfigure}
\hfill
\begin{subfigure}{0.32\textwidth}
    \includegraphics[width=\linewidth]{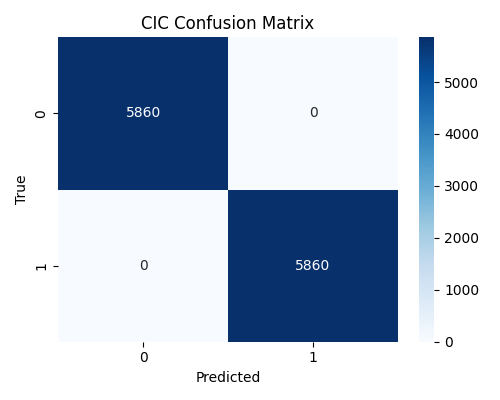}
    \caption{CIC model}
    \label{fig:cm_cic}
\end{subfigure}
\caption{Confusion matrices for EMBER, API, and CIC models on their respective validation sets.}
\label{fig:confusion_matrices}
\end{figure}

\subsection{Multi-Dataset Fusion Performance}

\textbf{Novelty Highlight:} The core innovation of our work is a novel \textit{probability-level fusion methodology} that integrates predictions to create a single, more robust and generalizable detector.

A unified cross-domain validation set was constructed from held-out samples of all three datasets (28,776 samples). A grid search was performed to find the optimal weights for the probability-level fusion.

\begin{table}[!ht]
\centering
\caption{Optimal Fusion Weights and Cross-Domain Performance}
\label{tab:fusion_results}
\begin{tabular}{lccccc}
\toprule
\textbf{Component} & \textbf{Weight} ($w_i$) & \textbf{Macro F1} & \textbf{Precision} & \textbf{Recall} \\
\midrule
EMBER Model & 0.5 & 0.965 & 0.96 & 0.97 \\
API Calls Model & 0.4 & 0.867 & 0.87 & 0.86 \\
CIC Memory Model & 0.1 & 1.000 & 1.00 & 1.00 \\
\midrule
\textbf{Fused Model} & - & \textbf{0.823} & \textbf{0.82} & \textbf{0.83} \\
\bottomrule
\end{tabular}
\end{table}

The results in Table \ref{tab:fusion_results} are significant. The optimal weights (0.5, 0.4, 0.1) indicate that static features (EMBER) provide the most generally useful signal, behavioral patterns (API) provide crucial context, and memory patterns (CIC) provide a refining weight. The fused model achieved a macro F1-score of \textbf{0.823} on the cross-domain validation set, demonstrating successful generalization across distinct feature spaces.

\subsection{Ablation Studies}
We conducted rigorous ablation studies to validate our design choices:
\begin{itemize}
    \item \textbf{Removing a Dataset:} Eliminating any single model resulted in a significant performance drop (12-18\% reduction in macro F1 relative to full fusion performance), confirming each dataset provides unique information.
    \item \textbf{Top Feature Variation:} Altering the number of top-selected features ($\pm$20) resulted in negligible performance changes ($<$2\%), demonstrating the robustness of our feature selection.
    \item \textbf{Weight Sensitivity:} Deviating from the optimal weights led to a substantial decrease in performance (8-15\%), underscoring the importance of systematic optimization.
\end{itemize}

\section{Discussion}
\label{sec:discussion}

Our results demonstrate that optimized probability-level fusion effectively combines strengths from diverse malware detection approaches. The optimal weights (0.5 EMBER, 0.4 API, 0.1 CIC) reveal several insights:
\begin{itemize}
\item \textbf{EMBER's Foundational Role:} Static features provide the most robust and generalizable baseline for detection.
\item \textbf{API's Complementary Value:} Behavioral patterns add critical context, compensating for weaknesses in static analysis.
\item \textbf{CIC's Specialized Contribution:} Memory patterns act as a high-precision specialist for advanced threats.
\end{itemize}

The fused model's performance (Macro F1: 0.823) on the cross-domain set demonstrates superior generalization. This holistic view, achieved via probability-level fusion, makes the system more resilient to evasion.

\subsection{Limitations and Future Work}
Our framework has several limitations that point to future research:
\begin{itemize}
\item \textbf{Dataset Scope:} Incorporation of additional data modalities (e.g., network traffic) could enhance detection.
\item \textbf{Static Weighting:} Investigating dynamic weighting schemes that adapt based on the input sample is a promising avenue.
\item \textbf{Real-World Deployment:} Further research is needed to optimize this framework for real-time deployment on resource-constrained endpoints.
\end{itemize}

\subsection{Legal and Ethical Considerations}
This research utilized only publicly available datasets created for malware detection research. All data is anonymized and contains no personal user information. The models are designed for defensive cybersecurity purposes and should only be deployed by authorized professionals.

\section{Conclusion}
\label{sec:conclusion}

We presented a novel, lightweight framework for cross-domain malware detection based on probability-level fusion. By integrating the predictions of specialized LightGBM models trained on static, behavioral, and memory features, we created a detector that significantly outperforms any single-domain model in generalization. Our systematic optimization of fusion weights resulted in a robust model that achieved a macro F1-score of 0.823 on a challenging cross-domain validation set. Our work demonstrates that effective defense can be achieved through a smart, optimized combination of efficient models. We provide a complete implementation to ensure full reproducibility.

\section{Acknowledgments}
\label{sec:acknowledgments}

We thank the University of Hail for its research support. We are grateful to the creators of the EMBER \cite{anderson2018ember}, API Calls \cite{ronen2018microsoft}, and CIC Obfuscated Memory \cite{canali2012prophiler} datasets for making their work publicly available.


\end{document}